# IoT-Enabled Low-Cost Fog Computing System with Online Machine Learning for Accurate and Low-Latency Heart Monitoring in Rural Healthcare Settings

Hamidreza Maneshti
Kermanshah, Iran
hamidreza.maneshti@gmail.com

Morteza Dadashi
*Dept. Computer Eng. & Information Tech Engineering*
Kermanshah, Iran
m.dadashi@razi.ac.ir

Kamyar Rostami
*Dep. Computer Science*
Montreal, Canada
kamyar.rostami.1@ens.etsmtl.ca

**Abstract:** Healthcare services in rural areas face numerous challenges due to the high cost of treatment and a lack of appropriate services. The application of Internet of Things (IoT) technology has shown potential in mitigating these issues. This article discusses the potential of Internet of Things (IoT) and fog computing to reduce healthcare costs and improve patient outcomes. The use of these technologies in cardiovascular health informatics is explored, along with the economic thought process of hospital decision-makers and end-of-life practices in intensive care units. Remote monitoring using IoT devices is highlighted as a promising way to detect health issues before they become serious, leading to earlier interventions and improved health outcomes. The use of fog computing in healthcare is also discussed, with a focus on its ability to provide real-time data processing, analysis, and decision-making capabilities. The article presents a novel architecture for Device-as-a-Service, utilizing both fog and cloud computing to improve the efficiency and accuracy of ECG device processing, and concludes that it has the potential to reduce costs by up to 80% in the Iranian market. The adoption of fog computing in healthcare is acknowledged to present significant challenges, such as security and privacy concerns, but its potential benefits make it a promising technology for the future of healthcare services.

Keywords: IoT, fog computing, heart monitoring, ECG, machine learning, rural healthcare, low cost, accuracy, reliability, online analysis, early detection.

Introduction:

Healthcare costs continue to rise, accounting for an estimated 10% of global GDP in 2019. In the United States, healthcare spending reached $3.8 trillion, or 17.7% of GDP in the same year. These escalating costs have prompted the need for innovative solutions to reduce expenses while improving the quality of care. The integration of the Internet of Things (IoT) and fog computing is a promising approach to achieving this goal. In particular, this approach can be applied to cardiovascular health informatics, where challenges and opportunities exist in the collection, storage, analysis, and dissemination of health data [1]. The economic thought process of hospital decision-makers is another crucial aspect to consider when exploring the potential of IoT and fog computing in reducing healthcare costs [2], [3]. By measuring their attitudes towards economic decisions, healthcare organizations can identify areas where IoT and fog computing can help reduce expenses. End-of-life practices in intensive care units are also important to consider [4]. By leveraging the power of IoT and fog computing, healthcare providers can optimize end-of-life care while minimizing costs.

In addition to reducing costs, IoT and fog computing have the potential to improve patient outcomes. For example, remote monitoring using IoT devices can help detect health issues before they become serious, leading to earlier interventions and improved health outcomes. According to a study by Grand View

Research, the global remote patient monitoring market is expected to reach $1.8 billion by 2026, driven in part by the increasing adoption of IoT in healthcare [5].

The application of fog computing in the healthcare industry has become increasingly significant due to its potential to enhance healthcare services through real-time data processing, analysis, and decision-making capabilities. With fog computing, vast amounts of data can be collected and processed from IoT devices in a distributed and efficient manner, enabling faster and more precise diagnosis and treatment. Remote patients can also benefit from personalized healthcare support, as evidenced by the fog-assisted personalized healthcare-support system for patients with diabetes [6]. Moreover, automated remote cloud-based heart rate variability monitoring systems have shown to be highly effective when using fog computing, as seen in [7].However, it is important to note that the adoption of fog computing in healthcare presents significant challenges, particularly in terms of security and privacy, as pointed out in [8]. Despite these challenges, the potential advantages of fog computing in healthcare make it a promising technology for the future of healthcare services [9], [10]. In summary, the use of IoT and fog computing in healthcare has the potential to significantly reduce healthcare costs, improve patient outcomes, and enhance the efficiency of the healthcare industry.

This paper presents an architecture for Device-as-a-Service, designed to implement an Internet of Things (IoT) based electrocardiogram (ECG) device. The architecture utilizes fog computing for pre-processing and cloud computing for online machine learning processing to detect various illnesses. After extensive testing of the ECG device, we conclude that it is highly reliable, and has the potential to reduce costs by up to 80% in the Iranian market. The proposed architecture for Device-as-a-Service is a novel approach that leverages the benefits of both fog and cloud computing to improve the efficiency and accuracy of ECG device processing. The use of machine learning algorithms for illness detection adds to the device's diagnostic capabilities, while the fog computing platform enables real-time data processing and reduces latency.

We conducted a trial of the device on a sample of four healthy individuals and compared its performance to that of a real ECG. Our analysis revealed that the device exhibited an error rate of 5.90%, but when compared to the normal value, it demonstrated an impressive accuracy rate of 98.17%. Additionally, to test the reliability of the device, we monitored its performance for a continuous two-hour period using Wireshark. Despite occasional internet disconnections, none of the data packets were lost, and the bandwidth remained steady at 5.80 k/s. These findings highlight the promising potential of this device for accurate and cost-effective ECG monitoring.

Literature review:

The article "A New Cloud and SOA Based Framework for E-Health Monitoring Using Wireless Biosensors" was authored by A Benharref and MA Serhani. The authors propose a new cloud-based framework for monitoring patient health using wireless biosensors. The idea is to use cloud computing and SOA (Service Oriented Architecture) principles to allow the framework to collect, store and analyze data from multiple wireless biosensors in real time.
In their experiments, the authors found that their framework can process large amounts of data from many sensors simultaneously. Best of all, it can handle up to 500 simultaneous sensor readings with a response time of less than 300 ms. Additionally, the framework was able to scale to support up to 5,000 sensor readings per second while maintaining response times below 500 ms. Overall, this article presents a promising approach to patient health monitoring using wireless biosensors. The authors propose that the use of a cloud-based framework will enable the collection and analysis of patient data in real time, enabling better monitoring and improved patient outcomes [11].

This question is Z. Rebolledo-Nandi、A. Chavez-Olivera、R. E. Caves-Valencia、A. Alarcon-Paredes and GA Alonso describe a new affordable and versatile mobile health monitoring system. The system collects data from sensors on the user's body and sends it to a central server for analysis using an Android app. The authors created the system using a microcontroller, a Bluetooth module and a smartphone. The author focused on developing an Android app that can connect to a microcontroller via Bluetooth and extract sensor data. We also designed a cloud-based system for storing and analyzing the data collected by the sensors. Their results showed that the system was highly accurate in measuring and transmitting data from various sensors such as blood pressure, heart rate and body temperature. Overall, this innovative, low-cost health monitoring system has potential for use in remote patient monitoring and other healthcare applications [12].

The article "Low-Power Wearable ECG Monitoring System for Remote Multi-Patient Monitoring" by E. Spanò, S. Di Pascoli, and G. Iannaccone discusses the development of a low-power wearable ECG monitoring system capable of remote multi-patient monitoring. I'm explaining. The authors used wireless communication systems and low-power design techniques to minimize system power consumption. The system has been tested on a group of healthy volunteers and shown to be effective in monitoring ECG signals with high accuracy and low power consumption. The system consumed only 1.2mW of power, had a sensitivity of 99.5%, and a positive predictive power of 98.8% [13].

An article by M. Ryan Fajar Nurdin, S. Hadyoso, and A. Rizal describes the development of a cost-effective Internet of Things (IoT) system for monitoring electrocardiograms (ECGs) in multiple patients. . The primary method is to use IoT devices that can simultaneously monitor the ECGs of multiple patients and wirelessly transmit the data to a central database. The system was tested on 10 patients, and the results showed that the system was effective in monitoring ECG signals and transmitting data accurately. the tests were done by Web Stress Tools, and benchmark apache. Moreover, used Apache Benchmark Tools Stress to simulate number of users. This article provides insight into the potential of his IoT system for low cost in healthcare and highlights the need for further research in this area [14].

B. Ramkumar, U. Satija, and M. S. Manikandan proposed a new method to detect changes in cardiac events during long-term healthcare surveillance. Their approach involves using statistical analysis and robust threshold-based techniques to identify significant heart rate variability and arrhythmias. The authors evaluated the technique using real patient data and found that he achieved 94.4% accuracy in detecting changes in cardiac events. This innovative method may be a promising tool for long-term monitoring of patients with cardiovascular disease [15].

Article by S.S. Naik et al. describes an intelligent Internet of Things (IoT) device for continuous patient monitoring in the intensive care unit (ICU). This device is used to monitor various physiological parameters of the patient such as body temperature, heart rate, blood pressure and oxygen saturation. The primary method used in this study is the development and implementation of IoT devices that can collect real-time patient data and send it to a central server. The device has been tested in a simulated ICU environment, and the results demonstrate its ability to accurately monitor and transmit patient data on an ongoing basis. The author concludes that IoT devices can be a valuable tool for improving patient care in his ICU [16].

The article "Real-Time Signal Quality Aware ECG Telemetry System for IoT-Based Healthcare Monitoring" by U. Satija, B. Ramkumar, and SM Manikandan presents a system for real-time monitoring of electrocardiogram (ECG) signals in health applications. It has been. of the Internet of Things (IoT). The authors developed an algorithm to analyze ECG signals and identify signal quality in real time. You can use it to detect and prevent potential health problems. The system was tested and achieved 93.7% accuracy in detecting signal quality. The authors conclude that the system has the potential to improve remote health monitoring and diagnosis of patients [17].

An article by J.H. Abawajy and MM Hassan proposes a Federated Internet of Things (IoT) and cloud computing system for comprehensive patient health monitoring. The proposed system uses a federation architecture that allows data to be shared between different medical institutions and devices. The authors report that their system achieves a high level of accuracy in patient health monitoring, achieving a mean absolute error of 1.3 beats per minute for heart rate monitoring and 0.6 degrees Celsius for body temperature monitoring. . The system also has a low latency of 0.2 seconds and a high throughput of 200 messages per second. The authors argue that their system could improve quality of care and reduce costs by enabling remote, real-time monitoring of patients [18].

The article via way of means of H. Zhang, J. Li, B. Wen, Y. Xun, and J. Liu, titled "Connecting smart matters in clever hospitals the usage of NB-IoT," The number one technique utilized in this newsletter is Narrowband Internet of Things (NB-IoT) generation, which connects smart matters in clever hospitals to the Internet of Things (IoT) community. The authors' intention is to enhance the performance of clever health facility systems, specially withinside the context of affected person tracking and information collection. The article affords effects withinside the shape of numbers, such as a assessment of the overall performance of NB-IoT and different IoT technologies, along with ZigBee and Wi-Fi. The authors file that NB-IoT has higher community insurance and penetration capabilities, in addition to decrease energy consumption, in comparison to different IoT technologies. They additionally offer facts at a success deployment of NB-IoT in numerous clever health facility scenarios, along with affected person tracking and system management. Overall, the thing shows that NB-IoT is a promising generation for connecting smart matters in clever hospitals, and has the capacity to seriously enhance the fine of healthcare services [19].

A.F. Hussein, A. Kumar and others. We introduce a remote heart rate variability monitoring system that uses a cloud-based approach. The system uses photoplethysmography (PPG) signals to calculate heart rate variability (HRV) and wirelessly transmits the data to the cloud for processing. The authors report that the system achieved 93.6% accuracy in detecting abnormal HRV patterns, demonstrating its potential for remote cardiovascular health monitoring [7].

M. Devarajan and et, describes a mist-assisted medical system for distant diabetic patients. The primary method used in this study is to use fog computing technology, which employs a network of distributed computing resources close to the patient, to develop a diabetics-focused environment rather than relying solely on centralized cloud computing. It was to develop a personalized healthcare support system. The system includes a wireless glucose sensor and an Android-based application to monitor glucose levels and provide personalized health advice. The authors report positive results from a small study in which 80% of participants were satisfied with the system and reported improved diabetes control. The study also showed a reduction in hospital visits and overall treatment costs [6].

Consequently, there is a lack of architectural designs that can provide affordable medical services such as ICU, MRI, and CT-Scans suitable for use in rural areas. In order to enhance this situation, we can utilize

Data as a Service (DaaS) for assessment and introduce heart monitoring devices that function as ECG devices and are IoT-based to monitor heart disease.

Methodology:

The proposed system, named "Device as a Service," has a general architecture as shown in Figure 1. The system aims to reduce costs by separating the calculation section from the data collection section. This separation allows for the use of cloud servers instead of powerful hardware for heavy processing tasks such as those required by CT-SCAN machines. The system also employs fog computing to preprocess the large amounts of data generated by sensors before transmitting it. Figure 1 depicts the proposed architecture, and the following sections will delve into the various layers of the architecture.

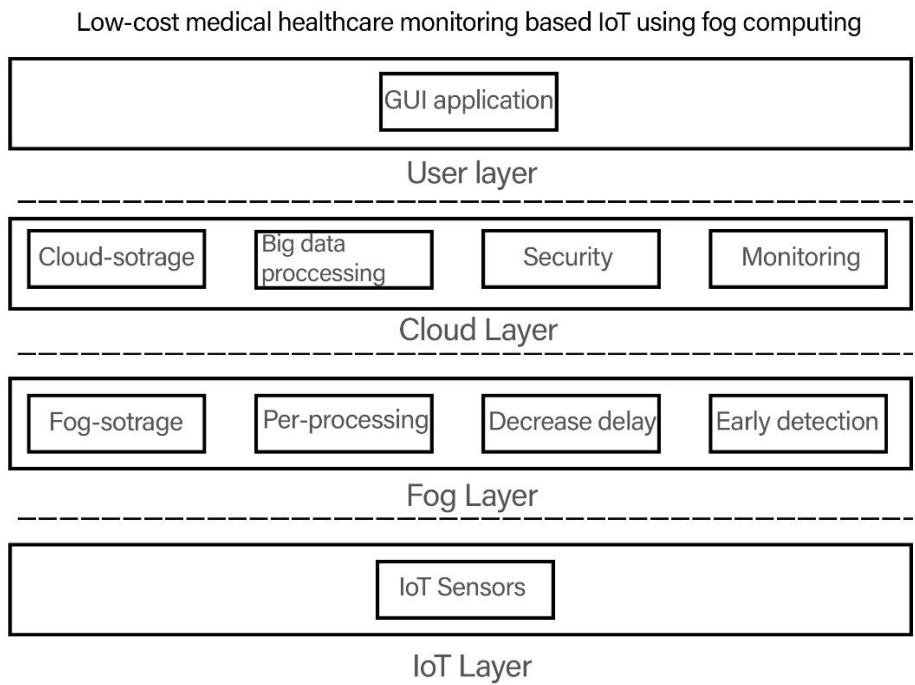

Figure 1: overall architecture Device-as-a-Service

Internet of things layer: The Internet of Things (IoT) layer is a crucial component of the system we're discussing. In this layer, we have a collection of sensors that are specifically designed to gather important biomedical data from humans. These sensors include things like ECG monitors, blood pressure sensors, body thermometers, and SpO2 sensors, to name a few. Once this data is collected, it is then converted from analog to digital format and sent up to the next layer of the system, which is known as the fog layer. Depending on the specific type of sensor being used, it may be connected to the fog layer either wirelessly or with a physical wire.

Fog layer: Within the fog layer of our system, digital data is not only stored temporarily, but also processed. This processing serves several purposes, including filtering the data to prevent unnecessary transmissions, minimizing latency in data transmission, and enabling quick identification of abnormal patient conditions, with immediate notification to the patient's caregivers. Furthermore, a user interface has been developed that utilizes a web-based application to allow patients to check the status of their device, as well as their own health status.

Cloud layer: The cloud layer plays a crucial role in our system, performing tasks such as processing large amounts of data, securely storing it, sending data for remote patient monitoring, and ensuring data security. Additionally, online machine learning techniques can be utilized to detect certain diseases, although this article focuses specifically on heart data. It is worth noting that additional data is required to diagnose diseases beyond the scope of heart data.

User layer: The user layer consists of a user interface that is provided as a mobile application and web-based application. Its main responsibility is remote patient monitoring.

Architecture of the proposed ECG device: Figure 2 represents the architecture of the ECG device based on the Device as a Service architecture. The layers of this architecture are essentially the same as the overall architecture, but there are some small differences, which we will discuss in more detail.

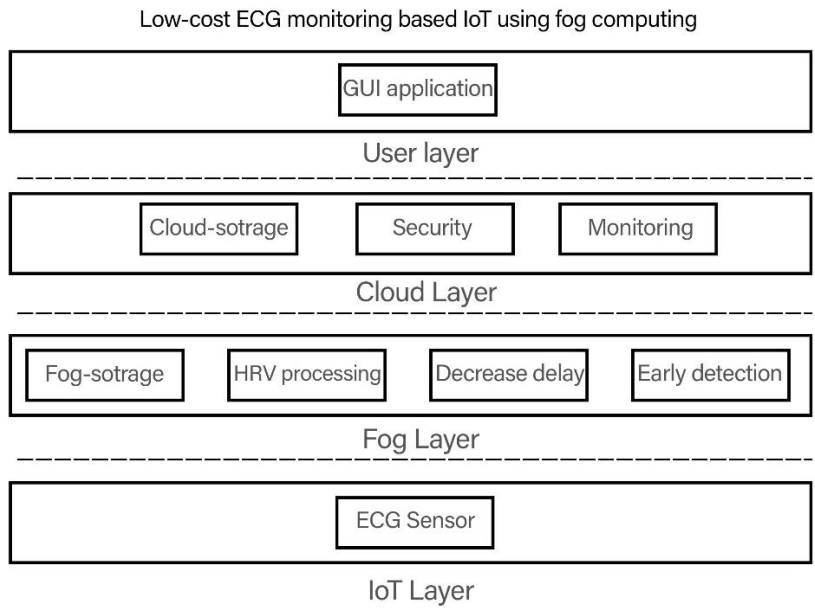

Figure 2: Architecture of the proposed ECG device

Proposed ECG device architecture (Figure 2): IoT layer: In the IoT layer of this architecture, the ECG sensor is connected to the fog layer through the USB2 port, and the received ECG signals from the patient are digitally converted and transferred to the fog layer.

Fog layer: The fog layer is similar to the general system architecture, with the difference that all processing takes place in this layer. The HRV algorithm processes the ECG signals to determine the changes in the patient's heart rate.

 Cloud server layer: In this layer, the only difference from the general architecture is the removal of the big data processing module since all processing takes place in the fog layer.

User layer: This layer is exactly the same as the general Device as a Service system architecture and has no differences.

Data flow in the proposed ECG system: Data flow in the proposed ECG system is as shown in Figure 3. The raw signals from the ECG sensor are sent to the fog node for HRV calculations and monitoring. The data is then sent to the cloud server via the internet for online monitoring. Finally, the processed signals and the calculated HRV values are transferred to the GUI for display to the user.

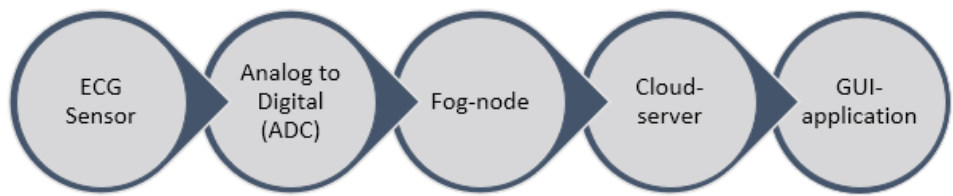

Figure 3: Data flow in the proposed ECG system

In this section, we presented a system architecture to help reduce healthcare costs. By implementing certain medical equipment using this architecture, it is possible to significantly reduce costs. We proposed an ECG device based on this architecture, which demonstrates that costs can be reduced using this approach. Overall, this system architecture can play an important role in reducing healthcare costs and making healthcare more accessible to a wider range of people.

Implementation design of the proposed ECG system: Figure 4 is a representation of the proposed system design and the modules used in this device. The implementation of the proposed system is both hardware and software. In future sections, we will examine both parts in detail.

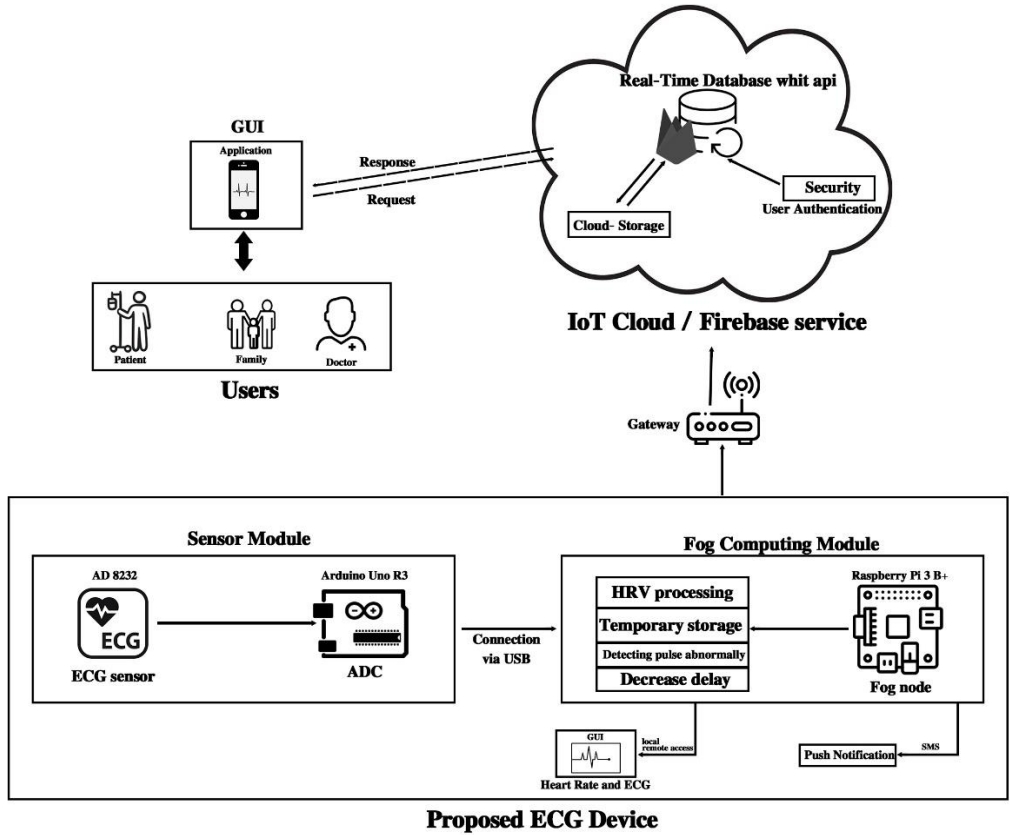

Figure 4: Implementation design of the proposed ECG system

To implement the hardware section, it is necessary to become familiar with the components of the proposed system, which are illustrated in Figure 5.

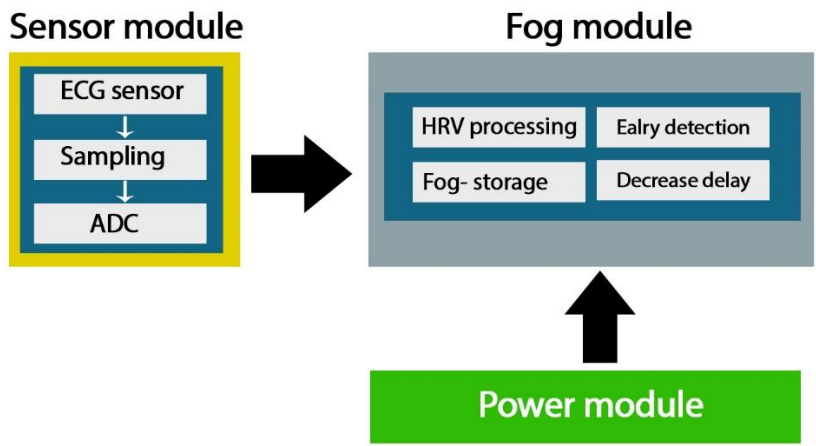

Figure 5: The main hardware components of the ECG monitoring system based on the Internet of Things

Sensor Module The sensor module consists of two main parts: the ECG sensor, which is responsible for collecting ECG signals, and the microcontroller (MCU) section, which includes data sampling and analog-to-digital conversion.

ECG Sensor: In this study, we used the AD8232 ECG module, which was considered important for two reasons: 1) its acceptable accuracy [20], and 2) its low cost, which is around 5,170,000 Iranian Rials (approximately 11 USD) in the global market.

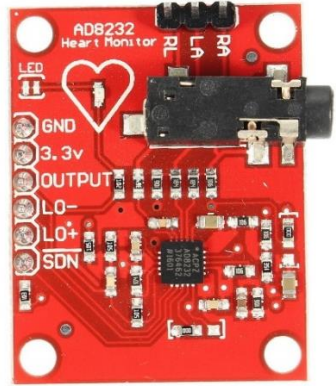

Figure 6: AD8232 model ECG module

As shown in Figure 6, there is a 3.5 mm headphone jack that is connected to three dry electrodes. Also, the location of the electrodes is shown in Figure 7, through which ECG signals are collected from the patient [21].

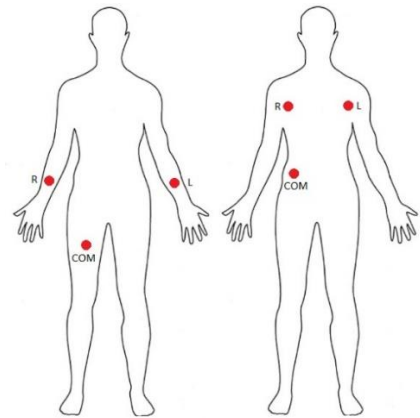

Figure 7: Placement of electrodes on the human body

According to the mentioned ECG sensor, the signals it produces are in analog form, so in the first step, they need to be sampled and then converted to digital data. To perform this process, we need a microcontroller that is capable of doing so. For this purpose, the Arduino Uno R3 is used, which is based on the ATmega328P microcontroller. It has high performance and low power consumption, as well as the ability to reduce noise during analog-to-digital conversion [22]. It is worth mentioning that the cost of this MCU is very low, with a price of about 9,400,000 Iranian Rials in the Iranian market and $20 in the global market.

In the MCU data processing section, data is divided into two stages: Stage 1: The sampling process takes place. Considering that the bandwidth of the ECG signal produced by the AD8232 module is usually between 50Hz and 100Hz [21]. According to the Nyquist theorem, the sampling rate on the Arduino board is set to 200Hz, so using a 16-bit timer, it creates a period for sampling and storage every 5 milliseconds. In addition, the timer interrupt priority is set to the highest possible level to prevent the effects of other interrupts such as external cuts or USART cuts. Figure 8 shows how the AD8232 module is connected to the MCU.

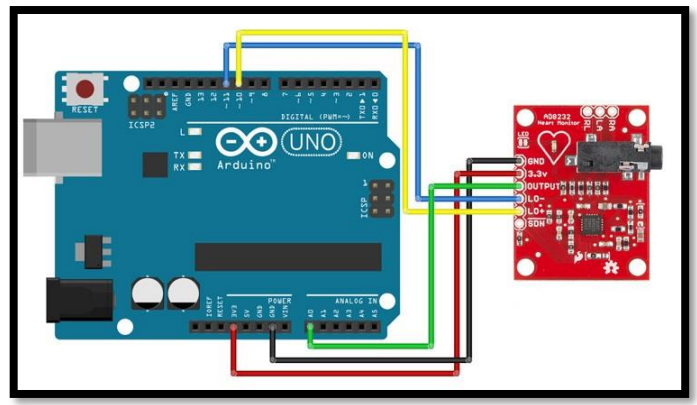

Figure 8: How to connect AD8232 to Arduino Uno R3

Stage two: The data conversion process takes place. The used MCU, thanks to ATmega328P, has 8 10-bit ADCs that perform analog-to-digital conversion in single-shot or scan modes.

Module Fog

In this section, the data converted by ADC or MCU is received in the fog node and prepared for processing. To this end, we used the Raspberry Pi 3 Model B+ (RSP) board with Raspbian jessie operating system. Furthermore, the MCU and RSP modules are connected via a USB 2.0 port with a baud rate of 115200 bps (i.e., 115200 bits are sent from the ADC to the RSP every second).

The Raspberry Pi 3 Model B+ board has received attention due to its low cost, which is about 20,000,000 Iranian Rials in the Iranian market and around $35 in the global market. Moreover, its hardware performance is suitable for the intended purpose [23].

The tasks of the fog module, in addition to being able to monitor the patient's condition on the patient side through the local network by a web-based program, are divided into 4 sections, including:

1. HRV processing: The task of the fog node or RSP board is to receive digital ECG data after analog to digital conversion through MCU. The received data is stored in RSP as a CSV file along with time stamp and counter label. In addition to displaying the heart rate waveform or ECG waves, which show samples of signals obtained by the device as shown in figures 9, HRV algorithm is used to calculate heart rate variability to simplify the analysis of the heart's condition.

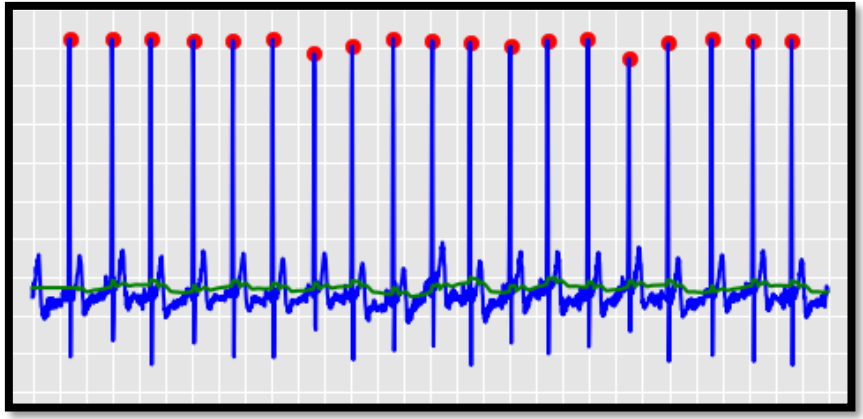

Figure 9: ECG produced by the presented device.

Early detection: After performing HRV calculations and identifying the patient's heart condition, it quickly informs via three methods, the internet, local network, and text message when the heart condition becomes abnormal.

Local storage: The proposed system is optimized for rural and underprivileged areas where the internet is usually available intermittently and sometimes with low speeds. For this purpose, local databases are used for temporary storage of data so that if the internet is disconnected during ECG device operation, the signals and results obtained from the patient's heart rate changes are not lost.

Delay reduction: To prevent the ECG waveform from being distorted, data is sent every 8 seconds along with a label in the order of processing, in case there is a delay in sending, the sent packets are not displaced, thereby preventing the ECG waveform from being distorted.

Power Supply :

Module Due to the low power consumption of its hardware components, the proposed system can be powered by either a 5-volt, 2.5-amp adapter or a recommended 6-volt, 4-amp battery.

Design and Implementation of the Software Component of the Proposed ECG Device

As previously stated, the proposed ECG device is used for two purposes, including generating ECG tracings and monitoring heart activity. Before discussing the software modules, we will briefly explain how the device operates in the two aforementioned modes, as shown in Figures 10 and 11. In this article, we provide a detailed description of the software modules used in the proposed ECG device. The modules are designed to acquire ECG data, process it, and display it on the device's screen. The device is equipped with various sensors and an AD converter to capture ECG signals. The signals are then filtered, amplified, and digitized before being processed using digital signal processing techniques.

The software component of the proposed ECG device has been developed using a variety of programming languages, including C, C++, and Python. We used C and C++ to write the code for the device's microcontroller and Python for the graphical user interface (GUI) and data processing modules.

The GUI module is responsible for displaying ECG waveforms and heart rate on the device's screen. It also provides various options for users to customize the display, such as zooming in and out of the waveforms and adjusting the heart rate calculation algorithm. In conclusion, the proposed ECG device is a reliable and efficient tool for monitoring heart activity and generating ECG tracings. The device's software component plays a crucial role in acquiring, processing, and displaying ECG data. The software modules have been designed and implemented using a variety of programming languages to provide a user-friendly GUI and accurate ECG data processing.

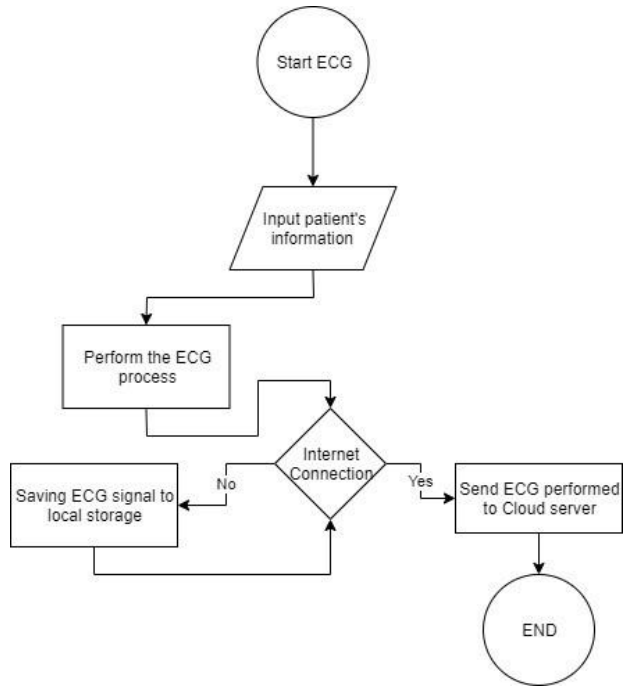

Figure 10: Flowchart of the implementation of the proposed system as an EKG device

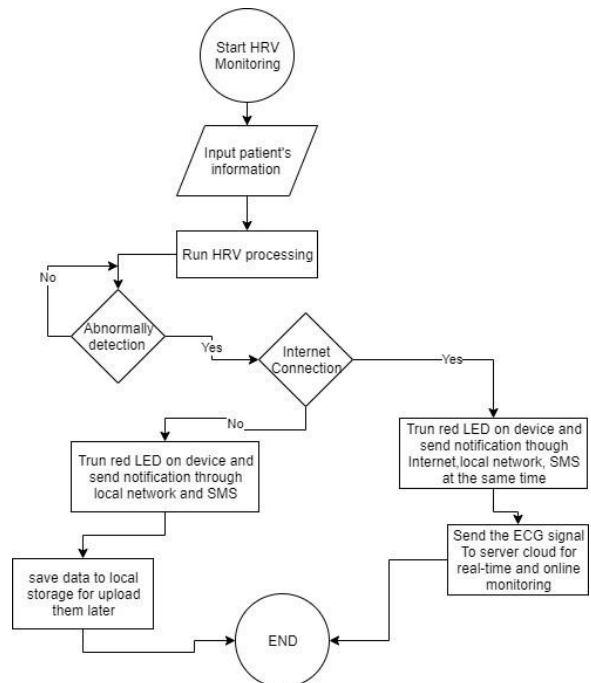

Figure 11: Flowchart of the implementation of the proposed system as a heart monitoring system

In order, to better understand the above steps, it is necessary to explain the components of the software modules in detail. The components of this section include several modules in the Python programming language, each of which has a specific responsibility. We will examine them in the following.

Module Processing The processing module is a collection of modules, each with a specific task. The tasks of these modules are as follows:

Data acquisition module: The purpose of this module is to receive raw signals from the analog-to-digital converter. To optimize processing time and reduce the number of writes to the database, the received signals are stored in separate CSV files, each containing 1500 signals. This module separates the signals into packets of 1500 and prepares them for processing. It should be noted that when storing signals in CSV files, a counter is saved as signal ordering and a time label.

Computation and storage management module: This module manages computations on the saved CSV files. It checks the CSV file and, if it has reached 1500 signals, executes the HRV computation module and sends the processed data along with the ECG signals to the cloud space. At the same time, it stores the signals and processed data in the cloud space and locally. The data is stored in the form of three lists containing 1500 ECG signals, time stamps, and counters, and a list containing the parameters processed by the HRV algorithm. This method prevents an increase in the number of writes to the database and optimizes data processing speed. This also justifies the use of cloud computing, as direct transmission of ECG data from the sensor to the cloud server would significantly increase the number of writes to the database, reduce data transmission speed, and increase the possibility of data loss. In the following sections, we discuss the use of local and cloud databases.

The proposed ECG device has two real-time databases, one based on the cloud and the other on the internet. These databases do not follow the traditional request-response protocols, which increases their efficiency.

In this article, we introduce and examine the two databases used:

Firebase Real-Time Database: In this research, we used Firebase Real-Time. The Firebase Real-Time Database is a cloud-based NoSQL database that enables the storage and use of data in applications through the internet. This research utilized the Real-Time Database service provided by Firebase, highlighting its exceptional data synchronization capabilities, cross-platform functionality, and high level of security. Data synchronization with the NoSQL cloud database ensures that data is updated and synchronized in real-time for all users, while the program's SDK enables data to be stored in internal memory when the application goes offline. This cloud web-based database stores data in JSON format, enabling synchronization for all connected users, regardless of platform or SDK.

One of the notable features of this database is its data synchronization protocol, which is faster and more efficient than HTTP, allowing devices to receive updated data in less than a millisecond. The high level of security provided by Firebase's authentication service helps to alleviate any concerns regarding data security[24], while read and write rules can be easily activated to further enhance security.

In conclusion, the Firebase Real-Time Database provides an effective and secure means of storing and synchronizing data in real-time for a wide range of applications, including those with different platforms and SDKs. This research highlights the benefits of using the Firebase Real-Time Database for data storage and synchronization, which is essential for modern applications that require real-time data exchange and efficient data management.

RethinkDB NoSQL database: RethinkDB is a JSON-based database that is open-source and free to use. It can be used locally without requiring an internet connection and can be installed on a personal server to provide online services. Instead of sampling the database to make changes, the database allows developers to perform real-time updates and view results instantaneously. This database is implemented based on the Real-Time push architecture [25].

As the device is used in rural and remote areas, optimization has been done to ensure that the data required by the doctor is not lost due to intermittent internet connections in these areas. Therefore, a database was used to store information offline. After connecting to the internet, the results of the processing are sent to Firebase. The reason for using a real-time database is to display the patient's status without requiring an internet connection on the patient's side.

Module for HRV calculations: The task of this module is to obtain Heart Rate Variability (HRV), which refers to the variation in the time interval between consecutive heartbeats, reflecting physiological signals that are equal to the rhythm of human heartbeats. HRV is an indicator of current cardiac irregularities and can serve as an early warning for imminent heart disease. These indicators are constantly present throughout the day or occur randomly[26]. This processing is carried out on the cloud server and is used to detect many complex heart diseases such as arrhythmia, myocardial ischemia, and Long QT syndrome using HRV signals[27], [28], which are analyzed by machine learning algorithms. To this end, we have used a simple and efficient algorithm that can implement the RSP metric. The algorithm is implemented in Python, and its initial version was proposed by Van Gent in 2016 [29]. In addition to being able to continuously analyze heart rate signals, this algorithm can filter signals and detecting errors when noise occurs.

In this research, we made modifications to this algorithm to enable it to detect vital signs during data collection. If vital signs occur, the device switches from offline to online mode and begins to send and store data. The algorithm consists of four sections, and we have customized the fourth section of this algorithm.

1. Part One: Identifying the First Peaks and Calculating the Heart Rate (BPM)
2. Part Two: Extracting the Complexity Value from the Heart Rate Signal
3. Part Three: Filtering the Signal and Improving Detection through Dynamic Thresholding
4. Part Four: Fast Detection and Device State Transition from Offline to Online

In Figure 12, we can see that this algorithm can effectively identify and separate noise from reliable data.

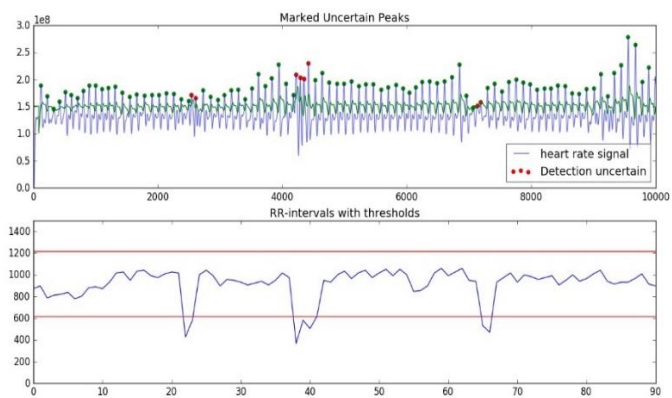

Figure 12: Sample signal with uncertain peaks

Results: To evaluate the experimental system, it is necessary to compare the proposed ECG device with real devices in terms of accuracy. However, since ECG signals are variable at any given time, it is not expected that the time values of the heart rate strips taken from both devices are equal. Nonetheless, the results obtained from both devices should be equal, indicating whether the patient's heart is healthy or not. To this end, the normal limits of these values have been expressed in articles [30], [31], such that if the parameters obtained from the heart rate strip are within this range, the heart is healthy. Time parameter (in meli seconds):

- Distance from RR 600 to 1200
- Distance from Q to T 320 to 440
- Distance from P to R 120 to 200
- Set QRS 80 to 100

The testing procedure was constrained by the inability to simultaneously connect the electrodes of both devices to the patient. Therefore, the experiment was conducted under identical conditions and immediately after conducting the test with the main ECG device, the test was repeated with the proposed ECG device. The test was performed on four individuals of varying genders and ages, with their physical characteristics listed in Table 1. The testing conditions for all four subjects were at rest and without any movement.

Table 1: Information of people

| Profile | Person Number one | Person Number two | Person Number three | Person Number four |
|---|---|---|---|---|
| Age | 20 | 24 | 47 | 53 |
| Gender | Female | Man | Female | Man |
| Heart status | health | health | health | Health |

Here, a condition applies: if the proposed ECG system produces results that indicate an unhealthy heart for an individual with a perfectly healthy heart, it indicates that the proposed system is not functioning correctly. It is worth mentioning that we used the SINA brand model 100 [32]electrocardiogram device, which is produced in Iran, for evaluation and comparison. Furthermore, in the conducted experiments, due to the proposed ECG device's ability to receive signals with three electrodes, we used the three electrodes of the reference device for the tests.

To obtain the accuracy of the proposed device, it is necessary to compare the obtained results with the reference values in terms of percentage errors. Then, the difference between the mean values of the waveform intervals of the proposed device and the main device is compared. The percentage of error is calculated using formula (1).

$$(1)\ \%error = \left|\frac{measured - accepted}{accepted}\right| * 100$$

The experiment was conducted empirically, in which two heart rate recordings were obtained from each individual - one with the ECG-SINA 100 machine and one with the proposed ECG machine. To validate and verify the results, it is necessary to obtain the time intervals of the ECG waves and then determine whether the heart is healthy or unhealthy. After obtaining the time intervals of the ECG waves in both results from the heart rate recordings, Tables 2 to 5 are formed for individuals one to four, respectively. These tables specify the time intervals of the ECG waves in milliseconds for four time periods.

Table2 : Results of error percentage number one

| Parameter | Normal values (in milliseconds) | ECG SINA 100 Device Values (in milliseconds) | Recommended ECG device values (in milliseconds) | Error percentage Relative to normal values (By Percentage) | Percentage error compared to main device (By Percentage) |
|---|---|---|---|---|---|
| RR-1 Distance | 600 to 1200 | 810 | 770 | 0 | 5.19 |
| QT-1 Distance | 320 to 440 | 332 | 320 | 0 | 3.75 |
| PR-1 Distance | 120 to 200 | 120 | 129 | 0 | 6.98 |
| QRS-1 Distance | 80 to 100 | 85 | 83 | 0 | 2.41 |
| Distance RR-2 | 600 to 1200 | 790 | 760 | 0 | 3.95 |
| QT-2 Distance | 320 to 440 | 335 | 317 | 0.93 | 5.68 |
| PR-2 Distance | 120 to 200 | 143 | 130 | 0 | 10.00 |
| QRS-2 Distance | 80 to 100 | 80 | 85 | 0 | 5.88 |
| Distance RR-3 | 600 to 1200 | 820 | 740 | 0 | 10.81 |
| QT-3 Distance | 320 to 440 | 345 | 325 | 0 | 6.15 |
| PR-3 Distance | 120 to 200 | 120 | 125 | 0 | 4.00 |
| QRS-3 Distance | 80 to 100 | 80 | 85 | 0 | 5.88 |
| Distance RR-4 | 600 to 1200 | 780 | 785 | 0 | 0.64 |
| QT-4 Distance | 320 to 440 | 320 | 320 | 0 | 0.00 |
| PR-4 Distance | 120 to 200 | 125 | 118 | 1.69 | 5.93 |
| QRS-4 Distance | 80 to 100 | 90 | 80 | 0 | 12.50 |
| | | | Total average | %1.31 | %5.61 |

Table 3 : Results of error percentage number Two

| Parameter | Normal values (in milliseconds) | ECG SINA 100 Device Values (in milliseconds) | Recommended ECG device values (in milliseconds) | Error percentage Relative to normal values (By Percentage) | Percentage error compared to main device (By Percentage) |
|---|---|---|---|---|---|
| RR-1 Distance | 600 to 1200 | 780 | 775 | 0 | 0.65 |
| QT-1 Distance | 320 to 440 | 350 | 330 | 0 | 6.06 |
| PR-1 Distance | 120 to 200 | 130 | 125 | 0 | 4.00 |
| QRS-1 Distance | 80 to 100 | 85 | 79 | 1.25 | 7.59 |
| Distance RR-2 | 600 to 1200 | 845 | 840 | 0 | 0.60 |
| QT-2 Distance | 320 to 440 | 365 | 355 | 0 | 2.82 |
| PR-2 Distance | 120 to 200 | 145 | 120 | 0 | 20.83 |
| QRS-2 Distance | 80 to 100 | 90 | 85 | 0 | 5.88 |
| Distance RR-3 | 600 to 1200 | 830 | 761 | 0 | 9.07 |
| QT-3 Distance | 320 to 440 | 320 | 345 | 0 | 7.25 |
| PR-3 Distance | 120 to 200 | 125 | 118 | 2.5 | 5.93 |
| QRS-3 Distance | 80 to 100 | 85 | 90 | 0 | 5.56 |

| Parameter | Normal values (in milliseconds) | ECG SINA 100 Device Values (in milliseconds) | Recommended ECG device values (in milliseconds) | Error percentage Relative to normal values (By Percentage) | Percentage error compared to main device (By Percentage) |
|---|---|---|---|---|---|
| Distance RR-4 | 600 to 1200 | 795 | 840 | 0 | 5.36 |
| QT-4 Distance | 320 to 440 | 320 | 330 | 0 | 3.03 |
| PR-4 Distance | 120 to 200 | 125 | 140 | 0 | 10.71 |
| QRS-4 Distance | 80 to 100 | 85 | 80 | 0 | 6.25 |
| | | | Total average | %1.87 | %6.35 |

Table 4: Results of error percentage number Three

| Parameter | Normal values (in milliseconds) | ECG SINA 100 Device Values (in milliseconds) | Recommended ECG device values (in milliseconds) | Error percentage Relative to normal values (By Percentage) | Percentage error compared to main device (By Percentage) |
|---|---|---|---|---|---|
| RR-1 Distance | 600 to 1200 | 735 | 725 | 0 | 1.38 |
| QT-1 Distance | 320 to 440 | 334 | 338 | 0 | 1.18 |
| PR-1 Distance | 120 to 200 | 153 | 190 | 0 | 19.47 |
| QRS-1 Distance | 80 to 100 | 85 | 80 | 0 | 6.25 |
| Distance RR-2 | 600 to 1200 | 760 | 680 | 0 | 11.76 |
| QT-2 Distance | 320 to 440 | 365 | 350 | 0 | 4.29 |
| PR-2 Distance | 120 to 200 | 130 | 122 | 0 | 6.56 |
| QRS-2 Distance | 80 to 100 | 83 | 85 | 0 | 2.35 |
| Distance RR-3 | 600 to 1200 | 720 | 675 | 0 | 6.67 |
| QT-3 Distance | 320 to 440 | 345 | 340 | 0 | 1.47 |
| PR-3 Distance | 120 to 200 | 135 | 123 | 0 | 9.76 |
| QRS-3 Distance | 80 to 100 | 80 | 80 | 0 | 0.00 |
| Distance RR-4 | 600 to 1200 | 715 | 690 | 0 | 3.62 |
| QT-4 Distance | 320 to 440 | 345 | 320 | 0 | 7.81 |
| PR-4 Distance | 120 to 200 | 125 | 120 | 0 | 4.17 |
| QRS-4 Distance | 80 to 100 | 85 | 81 | 0 | 4.94 |
| | | | Total average | %0 | %5.73 |

Table 5: Results of error percentage number Four

| Parameter | Normal values (in milliseconds) | ECG SINA 100 Device Values (In milliseconds) | Recommended ECG device values (in milliseconds) | Error percentage Relative to normal values (By Percentage) | Percentage error compared to main device (By Percentage) |
|---|---|---|---|---|---|
| RR-1 Distance | 600 to 1200 | 630 | 675 | 0 | 6.67 |
| QT-1 Distance | 320 to 440 | 330 | 345 | 0 | 4.35 |
| PR-1 Distance | 120 to 200 | 125 | 115 | 4.16 | 8.70 |
| QRS-1 Distance | 80 to 100 | 82 | 80 | 0 | 2.50 |
| Distance RR-2 | 600 to 1200 | 645 | 675 | 0 | 4.44 |
| QT-2 Distance | 320 to 440 | 335 | 350 | 0 | 4.29 |
| PR-2 Distance | 120 to 200 | 130 | 125 | 0 | 4.00 |
| QRS-2 Distance | 80 to 100 | 80 | 85 | 0 | 5.88 |
| Distance RR-3 | 600 to 1200 | 620 | 700 | 0 | 11.43 |
| $_{QT\text{-}3}$ *Distance* | 320 to 440 | 320 | 325 | 0 | 1.54 |
| $_{PR\text{-}3}$ *Distance* | 120 to 200 | 140 | 120 | 0 | 16.67 |
| $_{QRS\text{-}3}$ *Distance* | 80 to 100 | 80 | 80 | 0 | 0.00 |
| Distance RR-4 | 600 to 1200 | 680 | 715 | 0 | 4.90 |
| $_{QT\text{-}4}$ *Distance* | 320 to 440 | 345 | 340 | 0 | 1.47 |
| $_{PR\text{-}4}$ *Distance* | 120 to 200 | 135 | 120 | 0 | 12.50 |
| $_{QRS\text{-}4}$ *Distance* | 80 to 100 | 85 | 90 | 0 | 5.56 |
| | | | Total average | %4.16 | %5.93 |

Based on the results from tables 2 to 5, the ECG-SINA 100 device provided measurements within the normal range for each of the four individuals across all four time periods, indicating that they have healthy hearts. In contrast, the results from the proposed ECG device showed minor differences from the normal values for some time periods. However, these differences did not indicate any signs of an unhealthy heart for individuals one to four. We also calculated the percentage error of the results obtained from both the ECG-SINA 100 device and the proposed ECG device. It should be noted that due to variations in the electrical activity of the heart, the time intervals between the waves are naturally variable, leading to errors in the proposed ECG device compared to the ECG-SINA 100 device. The average error percentage for all four individuals in the proposed device is 5.90%, which indicates repeatability since the obtained values were close to each other. Therefore, this percentage difference is natural and has no effect on the final results, and as previously mentioned, both devices showed healthy hearts. The average error percentage relative to the normal values for all four individuals in the reference article was 1.83%, indicating an accuracy of 98.17% in the experiments conducted.

Technical and economic evaluation: The technical evaluation of the proposed system is assessed in three directions, and the evaluation of each of these three methods is described as follows:

Necessary bandwidth: In order to obtain the necessary bandwidth, we used the Wireshark software [33] of the proposed system. This software is used to determine the number of sent packets and their volume per unit time, and then by using formula (2), the system bandwidth is obtained.

$$(2)\ bandwidth = total\ packet\ size(bytes)/time\ span(s)$$

We send the ECG signal bandwidth along with the time stamp label and a list of parameters calculated by the HRV algorithm to the Firebase cloud server for evaluation. After approximately 2 hours of sending the data, the results of the necessary bandwidth and lost packets are shown in Figure 13, obtained from the Wireshark software. The information is presented in Table 6.

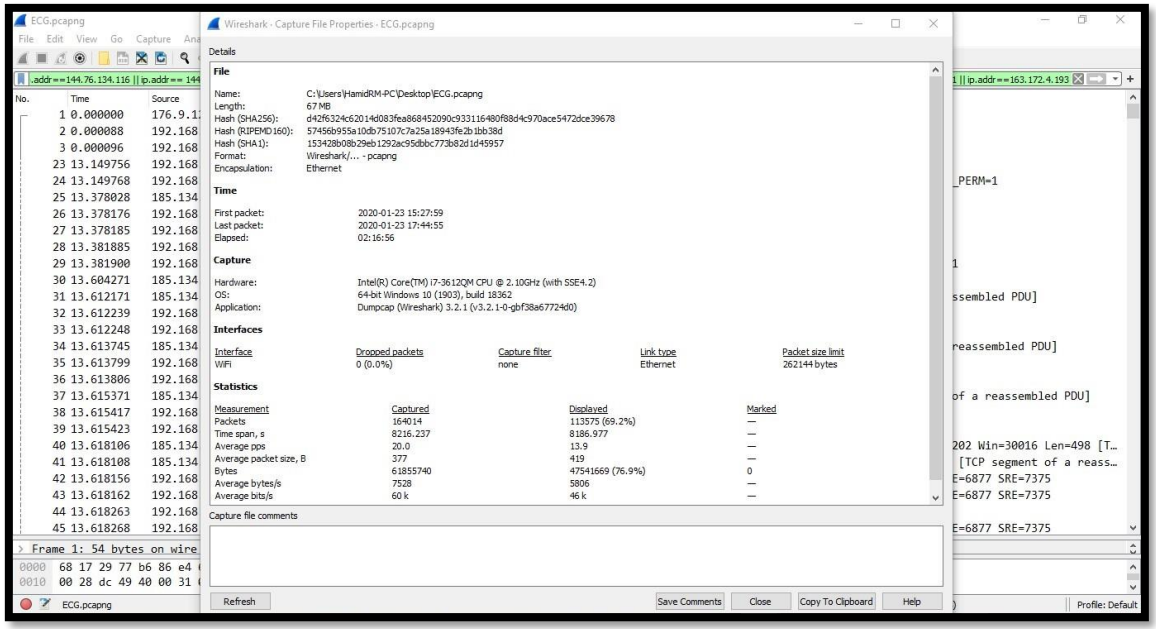

Figure 13: Wireshark software results

Table 6: The results of the packages sent and received.

| *Measured parameters* | *Values* |
|---|---|
| *Number of packages received and sent* | *113375* |
| *Total duration in seconds* | *8186.977* |
| *Average size of each package in bits* | *419* |
| *Total package size in bytes* | *47541669* |
| *Average bytes per second* | *5806* |
| *Average bits per second* | *46K* |
| *Missing Packages* | *0%* |

As a result, the obtained bandwidth from the above results is as follows:

*47541669/8186.977=5807 Byte/S = 5.80 Kilobyte/S*

Impact of Delay:

Due to the use of the proposed system in areas with a high probability of slow internet speeds, the delay in sending data packets increases, which affects the transmission of packets and may cause ECG waves to be displayed incorrectly. As mentioned in the previous section, to prevent this problem, signals are sent to cloud servers and local databases in lists of 1500. To test this section, we send 9 CSV files containing

1500 ECG signals, timestamps, and a list of 5 processed parameters by the HRV algorithm to the Firebase cloud server. Then, we compare the data before and after sending. Figures 14a and b show the data sent to the Firebase cloud server. As can be seen, there are 9 keys in Figure 14a, sorted in order of sending time, each with X and Y values. Therefore, Figure 14b shows that, for example, the Y values have been previously sent, which prevents delays in sending packets over the internet from affecting the order of data.

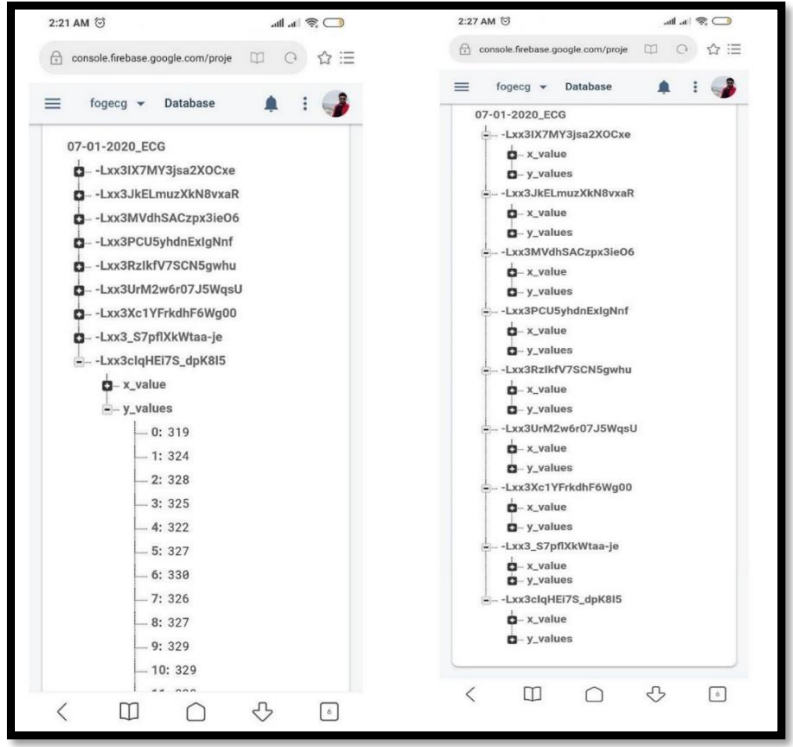

A                                    B

Figure 14: A view of the Firebase Cloud Server database

Power Consumption:

Based on the specifications of the board and modules used, we calculate the power consumption and energy consumption using the following formulas.

$$(3)\ w = P_{max} = \sum_{I=1}(V_i \cdot I_i)$$

$$(4)\ wh = P.T$$

In order to reach the maximum power and energy consumption of the proposed device, it is necessary to have the consumption of each of the boards and modules as mentioned in Table 7.

Table 7: Consumption rate of the proposed device by components

| Name and module | Voltage | Amp |
|---|---|---|
| Raspberry pi3 model b+[23] | 5.1 volt | 2.5 Ampere |
| *Brad Arduino Uno R3*[22] | *5 volt* | *20 mAmpere* |
| *MODULE ECG AD8232*[20] | *3.3 volt* | *20 microAmpere* |

As a result, the maximum power consumption of the proposed system is 12.85, and its maximum energy consumption in 24 hours is 308.4 watts per hour.

The proposed ECG device has been optimized for use and procurement in rural areas from two economic perspectives, with cost reduction being the main objective of this research. Therefore, it is necessary to compare the total cost of the entire device with similar cases. It should be noted that there is no similar commercial sample available in the Iranian market. In the global market, there are different examples, each with specific features, but our sample provides all the features, including support for the Internet of Things, simultaneous operation without the internet, and calculation of HRV in a single low-cost device. Table 8 shows a comparison with several devices available in the Iranian market, and as an economic cost optimization, cheaper boards such as Raspberry Pi Zero can be used.

Table 8: Comparing the system provided with similar devices inside the market.

| Name and model of the device | Device Application | key Features | Price in Toman | Country Builder |
|---|---|---|---|---|
| SINA 100 [32] | Electro cardio | Monitored<br>Storage capability of 150 ECG samples<br>Battery-powered<br>Weight about 3 kg<br>Unable to connect to the Internet | 16.500.00 | Iran |
| Carewell ECG- 1106L [34] | Electro cardio | 7 inch monitor<br>Ability to store 250 ECG samples<br>Ability to connect to computer from USB and LAN ports<br>ECG chart recording in 3-4-6 channel modes<br>Battery-powered<br>Weight about 3 kg<br>Unable to connect to the Internet | 17.700.000 | China |
| Kenz ecg-1210 [35] | Electro cardio | Pacemaker Detection<br>Monitored<br>Ability to store 100 ECG samples<br>Ability to transfer data to pc<br>Automatic measurement capability<br>Battery-powered<br>Weight about 3 kg<br>Unable to connect to the Internet | 28.320.000 | Japan |

| Proposed System [36] | Electro cardio Monitor heart rate changes | Ability to install monitors (increase the cost) <br> Unlimited ECG sample storage capability <br> Remote patient monitoring capability <br> Send notifications in abnormal heart condition <br> Usable through electricity and batteries <br> Usable in two offline and online fashions <br> Calculation of HRV parameters <br> Show warning go to device <br> Long-term usability | Raw price of the machine 5.0000.000 | Iran |
|---|---|---|---|---|

Conclusion:

The study proposes a Device as a Service (DaaS) architecture to provide medical equipment to healthcare centers in rural areas of Iran at a lower cost. An ECG device has been implemented based on this architecture to cater to the needs of healthcare centers in these regions. The system is designed to offer a heart monitor and ECG device at a lower cost than commercially available devices, enabling these centers to procure the equipment at a lower cost with better facilities. The proposed system has been compared with similar systems in the market, as shown in Table 5-1. The proposed device offers acceptable accuracy along with its cost-effectiveness, which has been obtained through experimental tests in the previous chapter. Additionally, the study suggests leveraging the Internet of Things (IoT) and machine learning for processing heart rate variability (HRV) changes, which can lower costs and reduce transportation risks from rural areas to nearby cities. Through this service, specialized physicians can monitor the heart condition of patients remotely, irrespective of time and location.

Recommendations Based on the experimental and analytical results obtained in this research, the following suggestions can be proposed for future work:

- Removing the device noise during patient movement.
- Implementing the CT-SCAN device using the proposed method and disease detection from CT-SCAN data.